\documentclass[a4paper,twocolumn,11pt]{quantumarticle}
\pdfoutput=1
\usepackage{lipsum}
\usepackage{amsmath}
\usepackage{amsthm}
\usepackage{amssymb}
\usepackage{graphicx}
\usepackage{esint}
\usepackage{titlesec}
\usepackage{amsfonts}
\usepackage{xcolor}
\usepackage[utf8]{inputenc}
\usepackage{amsfonts}
\usepackage{amsmath}
\usepackage{color}
\usepackage{footmisc}
\usepackage{bbold}
\usepackage{amsthm}
\usepackage{appendix}
\usepackage{graphicx}
\usepackage{halloweenmath} % also loads "amsmath" and "pict2e"
\usepackage{curve2e} % for polar coordinate in "picture" environments
\usepackage{hyperref}
\usepackage{bm}
\usepackage{float}

\usepackage{cancel}

\newtheorem{corollary}{Corollary}

\newtheorem{theorem}{Theorem}

\hypersetup{
    colorlinks=true,
    linkcolor=black,
    citecolor = black,
    filecolor=magenta,      
    urlcolor=cyan
    }

\makeatletter

\@ifdefinable\SuCmathpictvertex{} % check that this name can be used
\@ifdefinable\@SuC@reserved@dimen{\newdimen\@SuC@reserved@dimen}

\newenvironment*{@SuC@math@picture}[8]{%
  \def\SuCmathpictvertex{\circle*{#6}}%
  \setlength\unitlength{\fontdimen 22 #5\tw@}%
  \setlength\@SuC@reserved@dimen{#7\unitlength}%
  \kern\@SuC@reserved@dimen
  \@HwM@d@pict@strut{#2}%
  \picture(#3,#1)(#4,-1)%
    \roundcap
    \roundjoin
    \linethickness{#8\@HwM@thickness@units@for #5}%
}{%
  \endpicture
  \kern\@SuC@reserved@dimen
}
\newcommand*\@SuC@general@pict[9]{%
  \begin{@SuC@math@picture}%
            {#2}{#3}% height / depth
            {#4}{#5}% width / min-x
            #6% font selector (e.g., "\textfont")
            {#7}% diameter of the vertices
            {#8}% width of sidebearings
            {#9}% thickness of the edges
    #1%
  \end{@SuC@math@picture}%
}
\newcommand*\@SuC@math@version@shunt[7]{%
  \@HwM@choose@thicknesses{\@SuC@general@pict {#1}{#2}{#3}{#4}{#5}#7}%
      {{.4}{.2}{}}% for "normal" (actually, non-"bold") math version
      {{0.5}{.25}{0.75}}% for "bold" math version
}

\newcommand*\DeclareNewSuCMathPict[6]{%
  \newcommand*{#1}{%
    \@HwM@general@ordinary@symbol
      {\@SuC@math@version@shunt {#6}{#2}{#3}{#4}{#5}}%
  }%
}

\makeatother

\DeclareNewSuCMathPict{\pentagon}
            {3}{1}  % height / depth
            {4}{-2} % width  / min-x
{% code that draws the picture
    \polygon(90:2)(162:2)(234:2)(306:2)(378:2)%
    \put (90:2){\SuCmathpictvertex}%
    \put(162:2){\SuCmathpictvertex}%
    \put(234:2){\SuCmathpictvertex}%
    \put(306:2){\SuCmathpictvertex}%
    \put(378:2){\SuCmathpictvertex}%
}

\newcommand{\ket}[1]{\left| {#1}\right\rangle}
\newcommand{\bra}[1]{\left\langle {#1}\right|}
\newcommand{\braket}[2]{\left\langle {#1}\middle|{#2} \right\rangle}

\newcommand{\perm}[1]{\text{perm}\left({#1}\right)}

\theoremstyle{definition}
\newtheorem{conjecturep}{Conjecture P\ignorespaces}
\newtheorem{conjecturem}{Conjecture M\ignorespaces}
\newtheorem*{counter-example*}{counter-example}

\newtheorem*{theorem_stab}{Theorem}

\newcommand{\ad}{\hat{a}^{\dagger}}

\newcommand{\comm}[1]{\ignorespaces}

\begin{document}

\title{Anomalous bunching of nearly indistinguishable bosons}

%alternatice title: Enhanced bunching of nearly indistinguishable bosons

%alternative title: Enhanced boson bunching via near indistinguishability

%alternative title: Enhancing bunching via small perturbations to indistinguishable bosons

%alternative title: Enhancing boson bunching via small distinguishability perturbations 

%alternative title: Enhancing boson bunching via small perturbations to indistinguishable particles 

%alternative title: Enhancing bunching via small perturbations to indistinguishable bosons 

\author{Léo Pioge}
%\email{leo.pioge59@gmail.com}
\affiliation{Centre for Quantum Information and Communication, Ecole polytechnique de Bruxelles, CP 165/59, Universit\'e libre de Bruxelles, 1050 Brussels, Belgium}

\author{Benoit Seron}
%\email{benoitseron@gmail.com}
\affiliation{Centre for Quantum Information and Communication, Ecole polytechnique de Bruxelles, CP 165/59, Universit\'e libre de Bruxelles, 1050 Brussels, Belgium}

\author{Leonardo Novo}
%\email{leonardo.novo@inl.int}
\affiliation{International Iberian Nanotechnology Laboratory (INL), Av. Mestre José Veiga, 4715-330 Braga, Portugal}

\author{Nicolas J. Cerf}
%\email{nicolas.cerf@ulb.be}
\affiliation{Centre for Quantum Information and Communication, Ecole polytechnique de Bruxelles, CP 165/59, Universit\'e libre de Bruxelles, 1050 Brussels, Belgium}

\begin{abstract}
The commonly assumed straight link between boson bunching and particle indistinguishability in quantum interferometry has recently been challenged [Nat.~Photon. $\bm{17}$, 702 (2023)]. Exploiting the connection between quantum optical interferences and matrix permanents, it appeared that bunching effects may arise that exceed the expected limit of fully indistinguishable particles by injecting peculiar polarization states of partially distinguishable photons in some interferometers. Surprisingly, all states giving rise to such an anomalous bunching were found to be far from the state of fully indistinguishable particles, raising the question of whether this intriguing phenomenon might even possibly exist with \emph{nearly indistinguishable} particles. Here, we answer this question positively by relating it to a mathematical conjecture on matrix permanents dating from 1986, whose physical interpretation had not yet been unveiled. Using a recently found counterexample to this conjecture, we demonstrate that there is an optical interferometer involving 8~photons in 10~modes such that the probability that all photons bunch into two output modes can be enhanced by suitably perturbing the state of all photons having the same polarization. Such a finding reflects still another -- even less expected -- facet of anomalous boson bunching. 
\end{abstract}

\maketitle
\vspace{-5pt}

%%%%%%%%%%%%%%%%%%%%%%%%%%%%
\section{Introduction} \label{sec:Introdution}

In recent years, major strides have been made in the advancement of photonic quantum technologies,  driven by its wide-ranging applications in quantum communication, quantum computing, and quantum metrology \cite{pelucchi2022potential}. Central to the success of these applications lies the capability to generate, manipulate, and detect single photons, which has seen a remarkable progress. In particular, quantum photonics has led to one of the first claims of an experimental proof of quantum computational advantage over any classical algorithm \cite{zhong2020quantum} for solving the boson sampling problem \cite{aaronson2011computational}.  
Despite these remarkable developments, photonic quantum technologies are constrained by experimental limitations. In addition to the challenges posed by photon losses and various sources of noise, another constraint arises from the fact that photons possess various internal degrees of freedom, such as polarization, space, or time-frequency, and conventional photon sources do not yield ideal, perfectly indistinguishable photons. This difficulty prompts the need for an exploration of interferometry with partially distinguishable photons, halfway between fully distinguishable photons, governed by classical stochastic processes, and fully indistinguishable photons, where quantum superposition and bosonic statistics come into play \cite{tichy2015_partial_distinguishability, shchesnovich2015partial}.

Bosonic indistinguishability is at the heart of remarkable quantum interference phenomena, such as the celebrated Hong-Ou-Mandel (HOM) effect \cite{HOM}, which arises from the impossibility of distinguishing the situation in which two identical photons have crossed a 50:50 beam splitter from the trajectory in which they have both been reflected.
Destructive interference leads to the bunching of the two photons in the same output mode, an effect that becomes less pronounced as soon as the photons become partially distinguishable (for example, two orthogonally polarized photons do not bunch more than classical particles because they are perfectly distinguishable).  Larger scale photonic experiments involving many photons in many modes provide a testbed for the fundamental study of general boson bunching phenomena \cite{general_rules_bunching, carolan2014experimental, shchesnovich2016universality, young2023atomic}. The latter have been suggested to provide an efficient way to test the correct functioning of experiments which are difficult to simulate classically \cite{carolan2014experimental, shchesnovich2016universality}, relevant not only in photonics but also in atomic physics \cite{young2023atomic}.

A natural way to quantify bunching is to measure the probability that all photons coalesce in the same output mode of an interferometer. The probability of such single-mode boson bunching events can be seen as a measure of indistinguishability \cite{tichy2015_partial_distinguishability, general_rules_bunching}. However, in experiments involving many photons, this quantity is difficult to assess as this probability is exponentially small in the total photon number. This motivates the study of \textit{multimode} bunching probabilities, \textit{i.e.}, the probability that all photons coalesce in some chosen subset of output modes. These are easier to measure and, in most practical cases, their decrease witnesses again the fact that the particles become more distinguishable \cite{shchesnovich2016universality, seron2022efficient}. However, the behavior of multimode bunching probabilities with particle distinguishability is more subtle than one could naively expect. While numerical explorations and physical intuition strongly suggested that indistinguishable bosons should always maximize the bunching probabilities, it was recently demonstrated that this is not always the case. 
In Ref.~\cite{bosonbunching}, the authors presented a family of optical set-ups, involving 7 or more photons, in which partially distinguishable photons prepared in a specific polarization state can lead to a significantly higher multimode bunching probability than perfectly indistinguishable photons. The optical set-ups exhibiting this \emph{anomalous} boson bunching  were found by building upon 
a mathematical conjecture on matrix permanents stated by Bapat and Sunder in 1985~\cite{bapat1985majorization} and disproved by Drury more than thirty years later~\cite{drury2016}.

Although such instances tease our intuition about quantum interferences, a main question remained open in Ref.~\cite{bosonbunching} about the behavior of multimode bunching for input states in the vicinity of the state of perfectly indistinguishable photons. In fact, the peculiar states that were found to lead to anomalous bunching happened to be all located \emph{far} from the state of perfectly indistinguishable photons. Moreover, any perturbation to this state was shown to leave multimode bunching probabilities stationary (\textit{i.e.}, constant to the first order), making it very plausible that it is a local maximum.

In this paper, we investigate in depth the behavior of multimode bunching probabilities for \emph{nearly} indistinguishable bosons, revealing further counterintuitive phenomena and, in particular, contradicting the above assumption of a local maximum. First, we connect the question of whether boson bunching can only decrease following a perturbation of the state of indistinguishable bosons to a different conjecture on matrix permanents introduced by Bapat and Sunder in 1986 \cite{bapat1986extremal}, whose physical meaning had not yet been understood. Second, we convert a recent counterexample to this mathematical conjecture -- found amazingly again by Drury~\cite{drury2018} -- into a linear optical circuit, implying that multimode bunching can actually be \textit{enhanced} by applying a suitable perturbation to the photons' internal states. Our set-up requires 8 photons passing through a 10-mode interferometer, while the perturbation requires only the ability to manipulate a two-dimensional internal degree of freedom, such as polarization, making the experimental demonstration of this effect conceivable.

Finally, aiming at a better general understanding of the phenomenology of boson bunching, we investigate the perturbation that maximally decreases (instead of increasing) the bunching probability, starting again from the state of perfectly indistinguishable photons. We find out that perturbations of a two-dimensional degree of freedom, such as polarization, may actually decrease bunching probabilities even more than a perturbation towards the state of fully distinguishable particles, which necessarily involves at least an $n$-dimensional degree of freedom for $n$ photons (\textit{e.g.}, all photons occupying non-overlapping time bins). Although this feature is exhibited by single-mode as well as multimode bunching, so it is arguably less anomalous than enhanced bunching, it still defies intuition. We hope this work shall motivate further theoretical studies about the role of particle distinguishability in boson bunching, as well as experiments aimed at testing its anomalous behavior.

\section{Multimode boson bunching} 
\label{sec:Preliminary}
\label{sec:Multimode boson bunching}

In this section, we overview the quantum optics concepts that are needed to understand our work, starting with linear interferometry, partial distinguishability, and then proceeding with multimode boson bunching. Throughout this paper, we have a photonic platform in mind, where arbitrary interferometers are easily implementable. However, our results extend to any particles obeying bosonic statistics going through a linear interferometric process, which is also possible to implement with atomic experiments \cite{young2023atomic}. 

%%%%
\begin{figure}[t]
    \centering
    \includegraphics[width = 0.49\textwidth]{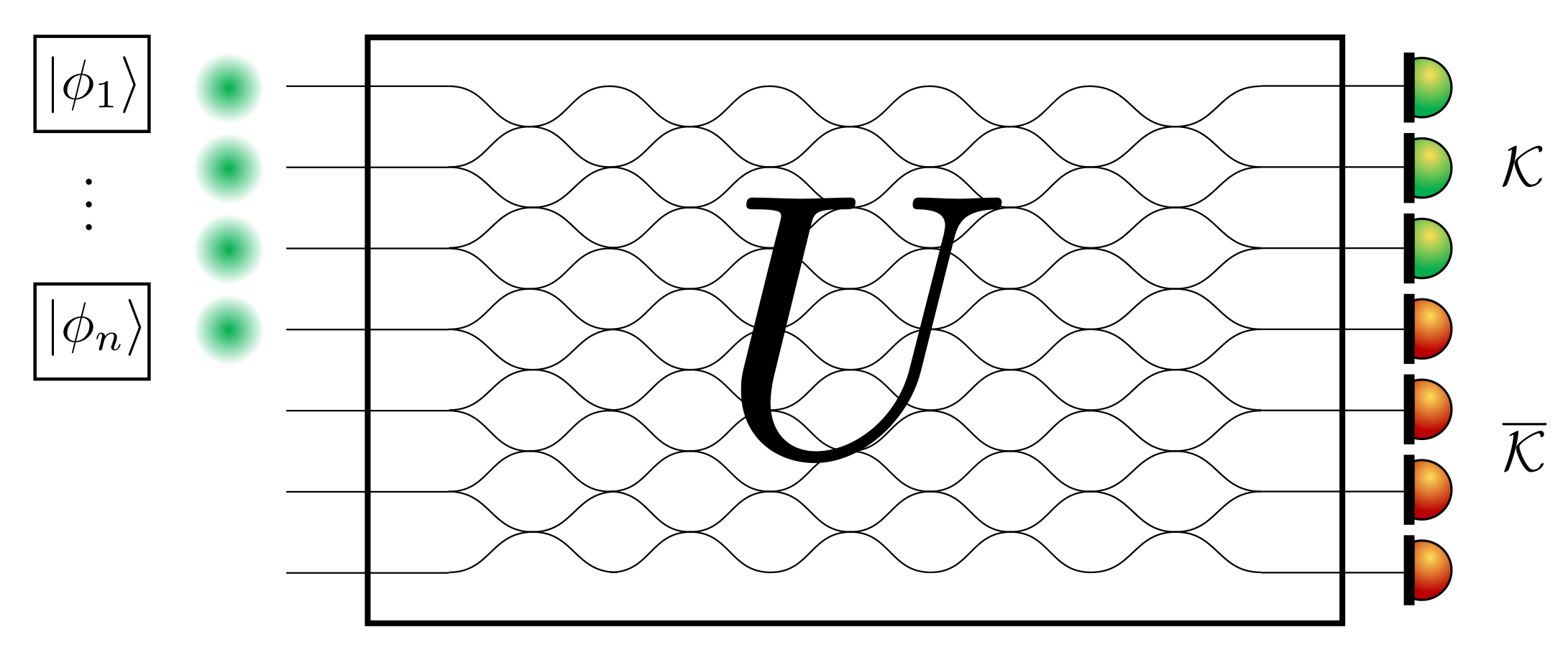}
    \caption{Interferometric setup where $n$ single photons enter the first $n$ spatial modes of a $m$-mode linear interferometer~$U$ (this interferometer can always be decomposed into a network of two-mode couplers). The photon entering the $j$th spatial mode carries some specific internal degrees of freedom (\textit{e.g.}, polarization, time, frequency) represented by the internal state $\ket{\phi_j}$.  
    We focus on the probability $P_n$ that all $n$~photons bunch into the subset $\mathcal{K}$ of output modes (indicated with green detectors) when the input photons are nearly indistinguishable, that is,  $\ket{\phi_j}\approx \ket{\phi_0}$.} 
\label{fig:generalInterferometer}
    %\vspace{1cm}
\end{figure}

Consider an $m \times m$ unitary matrix $U$, which represents an $m$-mode linear interferometer (see Fig.~\ref{fig:generalInterferometer}). We focus on a simple scenario where $n$ photons are sent into the first $n$ modes of this interferometer, with one photon per mode. In addition to being described by the spatial mode, the state of each impinging photon also depends on other -- called internal -- characteristics of the mode it occupies, such as its temporal or spectral distribution and its polarization. In what follows, the ket $\ket{\phi_j}$ will describe the internal state of the photon upon entering the $j$th spatial mode of the interferometer. Thus, as illustrated in Fig.~\ref{fig:generalInterferometer}, the input state is a pure, non-entangled state, namely  $\prod_{j=1}^n \ad_{j, \phi_j} \ket{0}$, where $\ad_{j, \phi_j}$ is the creation operator which creates a photon in the $j$th spatial mode with internal state $\ket{\phi_{j}}$. As commonly assumed, the interferometer keeps the internal state of the photons unchanged and only couples the spatial modes \cite{tichy2015_partial_distinguishability,dittel2018totally,shchesnovich2015partial}. Specifically, the interferometer that is associated with the unitary matrix $U$ (regarding its action on spatial modes) effects the unitary operator $\hat{U}$ defined as
\begin{equation}\label{eq:assumptionU}
    \hat{U} \, \ad_{j, \phi} \, \hat{U}^{\dagger}= \sum_{k=1}^m U_{k,j} \, \ad_{k, \phi},~~\forall j, \forall \phi,
\end{equation}
which does not act upon the internal degrees of freedom. For example, a single photon entering the $j$th spatial mode with internal state $\ket{\phi_j}$ exits the interferometer again in state $\ket{\phi_j}$ but in a superposition of spatial modes: it occupies the $k$th mode with probability amplitude $U_{k,j}=\bra{k}\hat{u}\ket{j}$, where $\hat{u}$ is the single-particle unitary operator while $\ket{j}$ and $\ket{k}$ denote, respectively, the input and output spatial state of the single photon.

The object of our study is the multimode bunching probability (also called generalized bunching probability \cite{shchesnovich2016universality, young2023atomic}), that is, the probability of finding all $n$ photons in a subset $\mathcal{K}$ of the output modes of the interferometer, see  Fig.~\ref{fig:generalInterferometer}. This probability, which we refer to as $P_n$, can be computed by projecting the output state $\hat{U} \prod_{j=1}^n \ad_{j, \phi_j} \ket{0}$
onto the subspace with all photons occupying only the output modes in $\mathcal{K}$. It can be shown to depend on the linear interferometer $U$ and subset $\mathcal{K}$ via the $n \times n$ positive semidefinite Hermitian matrix $H$, defined as
\begin{equation}
\label{eq:H_Shchesnovich}
H_{i,j} = \sum_{k \in \mathcal{K}} U_{k,i}^* U_{k,j}, \qquad i,j\in [n],
\end{equation}
with $[n] \equiv \{1,...,n\}$. Naturally, $P_n$ also depends on the $n \times n$ Gram matrix constructed with all possible overlaps between the internal states of the photons. 
The latter matrix, which is often referred to as the \emph{distinguishability} matrix and denoted as $S$ \cite{tichy2015_partial_distinguishability,shchesnovich2015partial}, is defined as
\begin{equation}
\label{eq_def_Gram_matrix_ij}
S_{i,j}=\braket{\phi_i}{\phi_j}, \qquad i,j\in  [n].
\end{equation}
 It appears that, for a given interferometer $U$ and subset $\mathcal{K}$, the multimode bunching probability depends on the distinguishability of the photons via \cite{shchesnovich2016universality}
\begin{equation}
\label{eq:bunching_probability}
P_{n}(S) = \perm{H \odot S},
\end{equation}
where $\perm{\cdot }$ denotes the matrix permanent while $\odot$ denotes the Hadamard (or element-wise) product, that is,  $(H \odot S)_{i,j} = H_{i,j} \,S_{i,j}$. This equation makes the connection between the physics of multimode boson bunching and the mathematics of matrix permanents.

%%%%%%%%%%%%%%%%%%%%%%%%%%%
\section{Physical implications of mathematical conjectures on matrix permanents}
%\subsection{Conjectures related to boson bunching}
\label{sec:Conjecture_&_Bunching}

\subsection{Global maximum of boson bunching}

Interestingly, several physical statements about boson bunching can be related to mathematical conjectures about the  permanent of positive semidefinite matrices which have remained open for years. Let us start with the following (very natural but false~\cite{bosonbunching}) physical conjecture about multimode boson bunching \cite{shchesnovich2016universality}.

\begin{figure*}[t]
    \centering
    \includegraphics[width = 0.78 \textwidth]{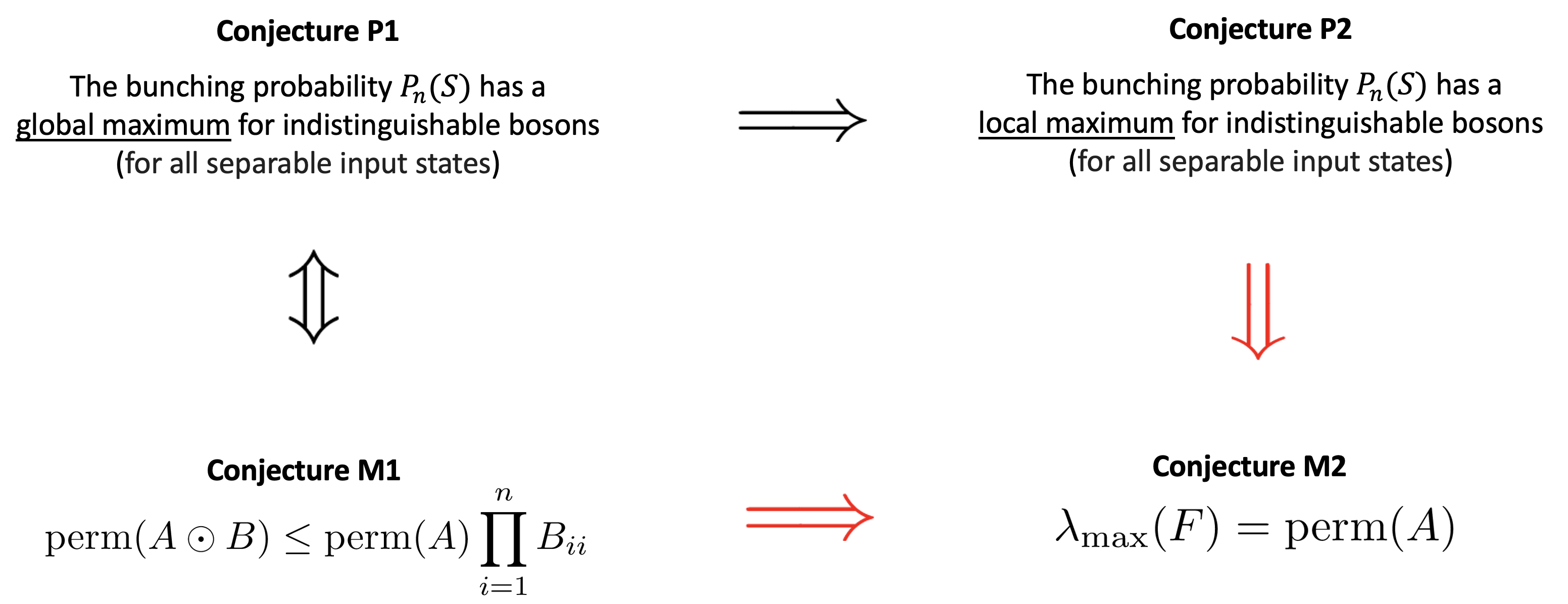}
    \caption{ Logical implications between the physical (labeled by P) and mathematical (labeled by M) conjectures discussed in Sec. \ref{sec:Conjecture_&_Bunching}. It had already been understood thatP\ref{conj:generalizedBunching} is equivalent to M\ref{conj:bapatSunder 1}\cite{shchesnovich2016universality}, which is why a counterexample to M\ref{conj:bapatSunder 1}could be exploited to find a counterexample toP\ref{conj:generalizedBunching} \cite{bosonbunching}. Our main contribution here  (represented by red arrows) is to introduce P2, a physical conjecture which is weaker than (hence implied by) P\ref{conj:generalizedBunching}, and show that its validity  would imply that  M\ref{conj:bapatSunder 2} is also valid. Consequently, we are able to use a counterexample to  M\ref{conj:bapatSunder 2} to construct a counterexample to P2. As a corollary, we also deduce a direct logical implication from M\ref{conj:bapatSunder 1}to  M\ref{conj:bapatSunder 2}, which is mathematical statement independent of the underlying physics \cite{math-paper}. } 
    \label{fig:conj_implication}
    %\vspace{1cm}
\end{figure*}

%%%%%
\begin{conjecturep}[Shchesnovich 2016]
\label{conj:generalizedBunching}
Consider any linear interferometer $U$ and any (nontrivial) subset $\mathcal{K}$ of output modes. Among all possible separable input states of $n$ classically correlated photons, the probability that all output photons bunch into $\mathcal{K}$ reaches its \emph{global} maximum if the photons are perfectly indistinguishable. 
\end{conjecturep}
%%%%
There was compelling evidence for the validity of this conjecture \cite{shchesnovich2016universality, bosonbunching}, which generalizes our intuition about the Hong-Ou-Mandel effect (namely, bunching is less pronounced as soon as the photons become partially distinguishable). Among other things, Conjecture~P\ref{conj:generalizedBunching}  holds for single-mode bunching  (when $|\mathcal{K}|=1$) since then the dependence of Eq.~\eqref{eq:bunching_probability} in $S$ factorizes, 
\begin{equation}
\label{eq:single-mode-factorization}
P_{n}(S) = \prod_{i=1}^n|U_{k,i}|^2 \,\perm{S} ,
\end{equation}
and $\perm{S}\le n!$. Furthermore, the opposite statement about fermion antibunching is true: a state of indistinguishable fermions always minimizes multimode bunching probabilities~\cite{shchesnovich2016universality}. Whereas this conjecture is valid for single-mode bunching, multiple bosons do not always comply with Conjecture~P\ref{conj:generalizedBunching}, as was surprisingly observed in Ref.~\cite{bosonbunching} via explicit counterexamples. Note that if we consider the particular case of a pure product input state as discussed in Sec.~\ref{sec:Multimode boson bunching}, where the photon in the $j$th spatial mode is in the pure internal state $\ket{\phi_j}$, Conjecture P\ref{conj:generalizedBunching} amounts to the following mathematical statement
\begin{equation}
\label{eq:conjectureP1}
   \perm{H \odot S} \leq \perm{H} .
\end{equation}
Here, we used Eq.~\eqref{eq:bunching_probability} as well as the fact that if all photons are indistinguishable, then $S = \mathbb{E}$ (with $\mathbb{E}_{i,j} = 1$ for all $i,j$) and so $\perm{H \odot \mathbb{E}}= \perm{H}$. In this form, it is clear that Conjecture P\ref{conj:generalizedBunching} is linked with the following mathematical conjecture 
introduced in 1985 in an article by Bapat and Sunder 
\cite{bapat1985majorization}.
\begin{conjecturem}[Bapat-Sunder 1985]
\label{conj:bapatSunder 1}
%Let $\mathbf{H_{n}}$ be the set of $n\times n$ positive semidefinite Hermitian matrices.
Let $A$ and $B$ be $n\times n$ positive semidefinite Hermitian matrices, then  
\begin{equation}
\label{eq:conj1}
    \perm {A\odot B} \leq \perm{A} \prod_{i=1}^n B_{ii}.
\end{equation}
\end{conjecturem}
It was shown in Ref. \cite{zhang1989notes} that, without loss of generality, we can restrict ourselves to the case where $B$ is a Gram matrix, that is, a Hermitian positive semidefinite matrix such that $B_{ii}=1$, $\forall i$. Thus, Conjecture M\ref{conj:bapatSunder 1} is equivalent to Eq.~\eqref{eq:conjectureP1}, which is itself equivalent to Conjecture P\ref{conj:generalizedBunching} restricted to pure product input states. Furthermore, since the bunching probability is linear in the input density operator, the existence of a counterexample to Conjecture P\ref{conj:generalizedBunching} with a mixed separable input state necessarily implies the existence of a counterexample for pure product input state(s). Therefore, Conjecture P\ref{conj:generalizedBunching} is simply equivalent to Conjecture M\ref{conj:bapatSunder 1} (see Fig. \ref{fig:conj_implication}).

In Ref.~\cite{bosonbunching}, the authors exploited a counterexample to Conjecture M\ref{conj:bapatSunder 1} found by Drury \cite{drury2016} to find several instances of interferometers and polarization states that contradict Conjecture P\ref{conj:generalizedBunching}. The simplest known instance involves $n=7$ photons with a well-chosen polarization state sent into a linear interferometer with $m=7$ modes and bunching into $|\mathcal{K}|=2$ output modes.
Even though all instances found in Ref.~\cite{bosonbunching} are such that $|\mathcal{K}|< n$, the multimode bunching conjecture is false for $|\mathcal{K}|\geq n$ too. Indeed, as shown in Appendix \ref{sec:appendix_counterexamples}, any counterexample to Conjecture P\ref{conj:generalizedBunching} with $|\mathcal{K}|< n$ can be converted to a physical situation where multimode bunching probabilities on subsets of size $|\mathcal{K}|\geq n$ can also be enhanced by a suitable state of partially distinguishable photons.

The focus of the present work lies in another central question that was left open in Ref.~\cite{bosonbunching}, namely whether the state of indistinguishable bosons remains at least a local (albeit not global) maximum for the multimode bunching probability.

\subsection{Local maximum of boson bunching}

It was noticed in Ref.~\cite{bosonbunching} that the distinguishability matrices $S$ realizing the known counterexamples to Conjecture P\ref{conj:generalizedBunching} are, in some sense, \emph{far} from the Gram matrix corresponding to indistinguishable photons, $S~=~\mathbb{E}$. This can be seen, for example, by using the measure of indistinguishability \cite{shchesnovich2015tight, tichy2015_partial_distinguishability, shchesnovich2016universality}   
\begin{equation} \label{eq:indistinguishability}
    d(S)=\frac{\perm{S}}{n!}, 
\end{equation}
which measures the weight of the permutationally symmetric component of the state $\ket{\Phi}~=~\ket{\phi_1}\ket{\phi_2}...\ket{\phi_n}$. This measure takes the value $d(\mathbb{E})=1$ for indistinguishable bosons, whereas it takes a value that is considerably smaller than $1$ for the internal photon states of the counterexamples from Ref.~\cite{bosonbunching}. This leads to a natural question: could multimode bunching probabilities be enhanced even for slightly distinguishable input states near the state of perfectly indistinguishable particles?  This question can be rephrased as follows. For a given interferometer $U$ and subset $\mathcal{K}$, we know after Ref.~\cite{bosonbunching} that the multimode bunching probability $P_n(S)$, seen as a function of the distinguishability matrix $S$, does not always have its global maximum for $S\equiv \mathbb{E}$. However, is this always a local maximum? The corresponding physical conjecture can be stated as follows.

\begin{conjecturep}[Local maximum] \label{conj:physical_conj}
Consider any linear interferometer $U$ and any (nontrivial) subset $\mathcal{K}$ of output modes. Starting from the state of $n$ indistinguishable photons, the probability that all output photons bunch into $\mathcal{K}$ can only decrease if a separable infinitesimal  perturbation is applied to the internal state of the photons, making them slightly distinguishable. 
\end{conjecturep}

Clearly, if Conjecture P\ref{conj:generalizedBunching} was true, then Conjecture~P\ref{conj:physical_conj} would also be true (see Fig. \ref{fig:conj_implication}). Since it is not the case, we need to bring another mathematical ingredient in order to address the validity of Conjecture P\ref{conj:physical_conj}. The main result of the present work is to reveal an unsuspected connection between Conjecture P\ref{conj:physical_conj} and a different mathematical  
conjecture which coincidently was also introduced by Bapat and Sunder, albeit in a different work \cite{bapat1986extremal}, and which funnily happens to have also been disproved by Drury~\cite{drury2018}. The conjecture is stated as follows. 

\begin{conjecturem}[Bapat-Sunder 1986]
\label{conj:bapatSunder 2}
Let $A$ be a $n\times n$ positive semidefinite Hermitian matrix. Let $F$ be the $n\times n$ matrix
defined as $F_{i,j}=A_{i,j} \, \perm{A(i,j)}$, where $A(i, j)$ denotes the $(n - 1)\times (n - 1)$ submatrix of $A$ obtained by deleting the $i$th row and $j$th column of $A$. Then, $\perm{A}$ is the largest eigenvalue of $F$.
\end{conjecturem}

Note that the matrix $F$ is positive semidefinite provided $A$ is positive semidefinite (see Appendix \ref{appendix:F_psd}), so all its eigenvalues are nonnegative. By using the Laplace expansion formula for the permanent, we have  $\sum_i F_{i,j}=\sum_j F_{i,j}=\perm A$. Thus, it is clear that $F$ always admits $\perm{A}$ as an eigenvalue (associated with the eigenvector with all equal entries). The conjecture is that $\perm{A}$ is its \emph{largest} eigenvalue. This can be proven to hold true for $n\leq 3$ and for the special case of nonnegative real matrices (i.e., such that $A_{i,j} \geq 0 $ for all $i,j$)  \cite{pate2008permanental}.  Incidentally, note also that $\det(A)$ can be proven to be the smallest eigenvalue of a matrix akin to $F$ but using determinants instead of permanents of submatrices.

The rest of this paper builds upon the connection between the mathematical Conjecture~M\ref{conj:bapatSunder 2} and the physical Conjecture P\ref{conj:physical_conj} on the behavior of boson bunching under small perturbations of the distinguishability matrix. Section \ref{sec:Small perturbation to indistinguishable boson} is devoted to the detailed perturbation analysis to the first order (Sec. \ref{sec:Stability around the bosonic case}) and second order (Sec. \ref{sec:second-order boson bunching}, where the link between these conjectures arises). Then, we show in Section~\ref{sec:counterexamples} how to construct a counterexample to Conjecture~P\ref{conj:physical_conj} from a counterexample to Conjecture M\ref{conj:bapatSunder 2}. Interestingly, this also implies a previously unknown logical implication between Conjectures M\ref{conj:bapatSunder 1} and M\ref{conj:bapatSunder 2}, which may be of independent interest to the mathematics community \cite{math-paper}.  The relation between the different physical and mathematical conjectures as well as our main new results are summarized in Fig.~\ref{fig:conj_implication}. 
Finally, we investigate the opposite question in Section~\ref{sec:Further analysis of multimode boson bunching}, namely what is the perturbation that maximally \textit{decreases} the bunching probabilities.

\section{Perturbations to indistinguishable bosons}
\label{sec:Small perturbation to indistinguishable boson}

Let us analyze how multimode bunching probabilities behave when small perturbations are applied to otherwise indistinguishable photons. As a reference situation, we consider that each of the $n$ input photons enters a different spatial mode of the interferometer and has initially the same internal state $\ket{\phi_{0}}$. We apply an infinitesimal perturbation to each photon's internal state, such that perturbation of the $i$th photon is proportional to an arbitrary (normalized) state $\ket{\eta_{i}}$. Specifically, the $i$th photon perturbed state  can be expressed as 
\begin{equation}
\label{eq:perturbed_bosonic_state}
     |\Tilde{\phi_{i}}\rangle=
     \frac{1}{\alpha_{i}}
     \Big(\ket{\phi_{0}} + \epsilon\,  v_{i} \ket{\eta_{i}}\Big),
\end{equation}
with $i\in [n]$, where $\epsilon>0$ stands for the perturbation strength and the $v_{i}$'s  are the components of an arbitrary complex vector $\bm{v}$ of constant norm. Without loss of generality we can choose $||\bm{v}||= 1 $, since this a rescaling of the components of $\bm{v}$ could be absorbed by a redefinition of $\epsilon$, i.e. if $\bm{v}\rightarrow \bm{v}/||\bm{v}||$ and $\epsilon\rightarrow ||\bm{v}|| \, \epsilon$. Further, we set 
\begin{equation}\label{eq:alpha_i}
\alpha_{i}=\sqrt{1+ 2\epsilon \, \mathfrak{Re}[v_i^*\braket{\eta_{i}}{\phi_{0}}] +\epsilon^{2} |v_{i}|^{2}}
\end{equation}
where $\mathfrak{Re}$ stands for the real part, so that the perturbed state $|\Tilde{\phi_{i}}\rangle$ remains normalized. Note that the most general separable perturbation to indistinguishable photons goes beyond Eq.~\eqref{eq:perturbed_bosonic_state}, but using the same argument as before (i.e., the probability is linear in the input density operator), we may restrict ourselves to perturbations in the form of a pure product state.

We start by proving a stationarity result, implying that first-order perturbations to the internal state do not change bunching probabilities (see Sec.~\ref{sec:Stability around the bosonic case}). This statement and a sketch of its proof had been given in Ref.~\cite{bosonbunching} (as it was expected that this stationary point would be a local maximum), but we present a more detailed proof here. The connection to Conjecture M\ref{conj:bapatSunder 2} appears from the second-order perturbation analysis (see Sec.~\ref{sec:second-order boson bunching}). As we shall see,
second-order perturbation terms may have a positive sign, so the state of indistinguishable photons is a stationary point that is \textit{not} a local maximum.

\subsection{First-order perturbation analysis}
%\subsection{Stability around the bosonic case}
\label{sec:Stability around the bosonic case}
Under a perturbation of the form given by Eq.~\eqref{eq:perturbed_bosonic_state}, the distinguishability matrix of the perturbed states becomes  
\begin{align}\label{eq:perturbed_S}
&\Tilde{S}_{i,j}=\braket{\Tilde\phi_{i}}{\Tilde{\phi_{j}}}\nonumber\\
   &=\frac{1+\epsilon\Big(v_{i}^{*} \braket{\eta_{i}}{\phi_{0}} +v_{j}\braket{\phi_{0}}{\eta_{j}}\Big) +\epsilon^{2}\, v_{i}^{*}v_{j}\braket{\eta_{i}}{\eta_{j}}}{\alpha_{i}\alpha_{j}}.
\end{align}
We will prove the following stationarity theorem.

\begin{theorem_stab} [Stationarity]
\label{th:bosonic_state_stable}
Consider any linear interferometer $U$ and any (nontrivial) subset $\mathcal{K}$ of output modes. For an  arbitrary choice of the perturbation parameters $v_{i}$ and $\ket{\eta_{i}}$ in Eq.~\eqref{eq:perturbed_bosonic_state}, the multimode bunching probability into $\mathcal{K}$ is stationary, i.e.,
    \begin{equation}
       \frac{\partial P_{n}(\tilde{S})}{\partial \epsilon} = 0.  
    \end{equation}
\end{theorem_stab}

\begin{proof}
    Recalling the definition of the normalization factors $\alpha_i$ from Eq.~\eqref{eq:alpha_i}, we can expand the perturbed distinguishability matrix $\tilde{S}$ from Eq.~\eqref{eq:perturbed_S} to first order as 
%%%%    
\begin{align}
\label{eq:gram_mat_first_order}
\Tilde{S}_{i,j}=1+i\, \epsilon \,  \mathfrak{Im}[v_i^*\braket{\eta_{i}}{\phi_{0}}+v_j\braket{\phi_{0}}{\eta_{j}}]+O(\epsilon^2), \nonumber \\
\end{align}
where $\mathfrak{Im}$ stands for the imaginary part. To simplify the notations, it is useful to define the vector $\bm{x}$ and the matrix $\delta S$ respectively as 
\begin{equation}
    x_i = \mathfrak{Im}[v_i^* \braket{\eta_{i}}{\phi_0}], \qquad   
\delta S_{i,j}=i\, \epsilon \,  (x_i-x_j), 
\end{equation}
so that the distinguishability matrix of the perturbed state can be reexpressed as
\begin{equation}
\label{eq:gram_mat_first_order_version_matrix}
     \Tilde{S}= \mathbb{E}+\delta S +O(\epsilon^2).
\end{equation}
According to Eq.~\eqref{eq:bunching_probability}, for a fixed interferometer $U$ and subset $\mathcal{K}$, the multimode bunching probability of the perturbed state can be expressed as
\begin{align}
\label{eq:bunching_probability_perturb}
      P_n(\Tilde{S})&=\perm{H\odot \Tilde{S}}  ~~~~~~~~~~\nonumber \\
     &=\perm{H\odot\mathbb{E}+H\odot \delta S } +O(\epsilon^2)\nonumber \\
\end{align}
By using the Minc's formula [Eq.~\eqref{eq:minc} from Appendix~\ref{sec:Minc'sformula}] to write the permanent of a sum of matrices, the multimode bunching probability becomes 
\begin{align}
     P_n(\Tilde{S})&= \perm{H} \nonumber \\
     &+ \underbrace {\sum_{i,j}^{n}(H\odot \delta S)_{i,j}\, \perm{H(i,j)}}_{O(\epsilon)}+O(\epsilon^2).
\end{align}
%where we have used Minc's formula [Eq.~\eqref{eq:minc} from Appendix~\ref{sec:Minc'sformula}] to write the permanent of a sum of matrices. 
%By applying this formula to Eq.~\eqref{eq:bunching_probability_perturb}, we derive
%\begin{align}\label{eq:expansion_bunching}   &\perm{H\odot \Tilde{S}^{T}}= \perm{H}   \nonumber\\      &+ \epsilon \sum_{i,j}^{n}(H\odot \delta S^{T})_{i,j}\perm{H(i,j)}+O(\epsilon^2).   \end{align} 
For notational brevity, we define the matrix $F$ as 
\begin{equation}
\label{eq:F_definition}
  F_{i,j}=H_{i,j}\, \perm{H(i,j)},
\end{equation}
which allows us to write the linear perturbation term of Eq.~\eqref{eq:bunching_probability_perturb} in a simpler form, namely
\begin{equation}
     \sum_{i,j}^{n}(F\odot \delta S)_{i,j}=i\, \epsilon \sum_{i,j}^{n} (x_{i}-x_{j} )F_{i,j}.
\end{equation}
Remarkably, the matrix $F$ appearing in Conjecture M\ref{conj:bapatSunder 2} appears naturally in the perturbative expansion of the bunching probability. As already mentioned, the Laplace expansion formula for permanents implies that, for any choice of a row or column of $F$, 
\begin{equation}\label{eq:sumrowcolumnF}
  \sum_{i=1}^{n} F_{i,j} =\sum_{j=1}^{n} F_{i,j}=\perm{H}.   
\end{equation} 
Hence, 
\begin{align}
      \sum_{i,j}^{n} (x_{i}-x_{j} )F_{i,j} \nonumber 
      &= \left(\sum_{i}^{n} x_{i}  - \sum_{j}^{n} x_{j}\right) \, \perm{H}\\ 
      &= 0, 
\end{align}
    which concludes the proof. 
\end{proof}

The stationarity of the multimode bunching probability near fully indistinguishable photons supports the idea that, even though we now know it is not a global maximum~\cite{bosonbunching}, it might be a local maximum (Conjecture P\ref{conj:physical_conj}). In what follows, we show that the second-order expansion of the bunching probability invalidates this conjecture.

\subsection{Second-order perturbation analysis}
\label{sec:second-order boson bunching}
In the second-order expansion, we will assume that the perturbation of each photon in Eq.~\eqref{eq:perturbed_bosonic_state} is orthogonal to the initial internal state $\ket{\phi_0}$, that is, 
\begin{equation}\label{eq:perturbations}
\braket{\phi_0}{\eta_i}= 0, \qquad i\in  [n].
\end{equation}
This simplifies the analysis and is actually sufficient to connect the variation of the multimode bunching probability to Conjecture M\ref{conj:bapatSunder 2}. Using this assumption, 
%normalization factor simplifies $\alpha_{i}=\sqrt{1+\epsilon^{2} |v_{i}|^{2}}$ and 
the perturbed distinguishability matrix can be written as
\begin{align}
    \Tilde{S}_{i,j}=\braket{\Tilde{\phi_{i}}}{\Tilde{\phi_{j}}}=
    \frac{1+\epsilon^{2}v_{i}^{*}v_{j}\Delta_{i,j}}{\sqrt{(1+\epsilon^{2} |v_{i}|^{2})(1+\epsilon^{2} |v_{j}|^{2})}},\nonumber \\ 
\end{align}
where we defined the Gram matrix associated with the perturbation $\Delta_{i,j}= \braket{\eta_i}{\eta_j}$. By applying Taylor's expansion and neglecting the terms of $\mathcal{O}(\epsilon^{4})$, we obtain
\begin{align}
\Tilde{S}_{i,j}=1+\epsilon^{2}\Bigl(v_{i}^{*}v_{j}\Delta_{i,j}-
    \frac{|v_{i}|^{2}}{2}-\frac{|v_{j}|^{2}}{2} \Bigl)+\mathcal{O}(\epsilon^{4}).
\end{align}
Note the absence of linear terms in $\epsilon$ in $\Tilde{S}_{i,j}$ due to the fact that $\braket{\phi_0}{\eta_i}= 0$. By defining the matrix $X$ such that $\Tilde{S}_{i,j} = 1 + \epsilon^{2}X_{i,j} +O(\epsilon^4)$, the probability of multimode bunching for the perturbed state is equal to
\begin{align}
     P_{n}(\Tilde{S})&=\perm{H \odot \Tilde{S}} \nonumber \\
     &= \perm{H \odot\mathbb{E} +\epsilon^{2}H \odot X} + O(\epsilon^4).
\end{align}
Using again Minc's formula (see Appendix \ref{sec:Minc'sformula}) as well as the definition of matrix $F$ [Eq.~\eqref{eq:F_definition}], we obtain 
\begin{align}
\label{eq:second_order_bunching}
    P_{n}(\Tilde{S}) = \perm{H}+\epsilon^{2} \sum_{i,j}^{n}(F\odot X)_{i,j}+O(\epsilon^4).
\end{align}
We note that this equation has no linear terms in $\epsilon$, which is consistent with the stationarity theorem. By plugging in the definition of $X_{i,j}$, we obtain the variation of the bunching probability, namely
\begin{align}
\label{eq:second_order_bunching_with_vi}
     &\delta P_n= \perm{H \odot \Tilde{S}} - \perm{H} \nonumber\\
     &\ = \epsilon^{2} \sum_{i,j}^{n}\Big(v_{i}^{*}v_{j}\Delta_{i,j}-\frac{|v_{i}|^{2}}{2}-\frac{|v_{j}|^{2}}{2}\Big)F_{i,j}+O(\epsilon^4) \nonumber \\
     &\ = \epsilon^{2} \Bigl( \bm{v}^{\dagger} (\Delta \odot F)\,  \bm{v}
     - \perm{H}\Bigl)+O(\epsilon^4).
\end{align}
Our aim is to determine whether $\delta P_n$ could possibly be positive, which would mean that the bunching probabilities are not always locally maximized by the state of indistinguishable photons. To do so, we first demonstrate an upper bound on $\delta P_n$ related to the maximum eigenvalue of $F$ and subsequently show how to saturate this upper bound. Since $\Delta$ is a positive semidefinite (Gram) matrix, it admits a decomposition 
$\Delta=M^{\dagger}M$, where $M$ is a $K\times n$ matrix and $K$ is the rank of $\Delta$. Specifically,
\begin{equation}
\Delta_{i, j} = \sum_{k=1}^{K} (u_{i}^{k})^{*} \, u_{j}^{k},
\end{equation}
where  $\bm{u}^{k}$ with $k=1,\cdots K$ denotes a set of $K$ linearly-independent $n$-dimensional column vectors whose components are $u_{i}^{k}$, where $\sum_{k=1}^K |u_{i}^{k}|^2 = \Delta_{i,i}=1$, $\forall i$ \cite{gloub1996matrix}. 
Using this fact together with Eq.~\eqref{eq:sumrowcolumnF} and $||\bm{v}||=1$, we can rewrite Eq.~\eqref{eq:second_order_bunching_with_vi} as  
\begin{align}
    \label{eq:before_upper_bound}
     \delta P_n=&\epsilon^{2} \Bigl( \sum_{k=1}^{K}(\bm{v}\odot \bm{u}^{k})^{\dagger} F (\bm{v}\odot \bm{u}^{k})
     - \perm{H}\Bigl) \nonumber\\
     &+O(\epsilon^4).
\end{align}
We can upper bound the right-hand side term of this equation by using 
\begin{equation}        
\sum_{k=1}^{K}(\bm{v}\odot \bm{u}^{k})^{\dagger} F (\bm{v}\odot \bm{u}^{k}) \leq ||F|| \,
    \sum_{k=1}^{K}||\bm{v}\odot \bm{u}^{k}||^{2},
\end{equation}
which can be further simplified by noting that 
\begin{align}
\label{eq:norm_v_o_u_equal_1}
    \sum_{k=1}^K ||(\bm{v} \odot \bm{u}^{k})||^{2}&=\sum_{k=1}^K \sum_i^n|v_i|^{2}| u^{k}_i|^{2}\nonumber  \\& =\sum_{i=1}^n|v_i|^{2}\Delta_{i,i}=1.
\end{align}
Since the matrix norm $||F||=\lambda_{\max}(F)$ is given by the largest eigenvalue of $F$, we obtain the upper bound 
\begin{align}  
    \label{eq:upper_bound}
    \delta P_n\leq \epsilon^2 \Big(\lambda_{max}(F)-\perm{H}\Big)+O(\epsilon^4).
\end{align}
In order to show how to saturate this upper bound, we start by  choosing all states $\ket{\eta_i}\equiv \ket{\phi_1}$ with $\braket{\phi_0}{\phi_1}=0$, which is equivalent to choosing the Gram matrix $\Delta= \mathbb{E}$ for the perturbation. Note that the perturbed internal states $|\Tilde{\phi_{i}}\rangle \propto\ket{\phi_{0}} + \epsilon\,  v_{i} \ket{\phi_{1}}$ span only a 2-dimensional space here.  The matrix $\Delta$ is of rank one, hence its decomposition has a single vector $\bm{u}^1= (1, 1,...,1)^T$.  Plugging this into Eq.~\eqref{eq:before_upper_bound}, we obtain 
\begin{align}\label{eq:maximum_bunching}
    \delta P_n = \epsilon^{2}  \Bigl(\bm{v}^{\dagger} F \bm{v} - \perm{H}\Bigl) +O(\epsilon^4).
\end{align}
It is thus possible to saturate  the upper bound of Eq.~\eqref{eq:upper_bound} by choosing $\bm{v}$ as the eigenvector $\bm{v}_{\max}(F)$ of $F$ corresponding to its largest eigenvalue $\lambda_{max}(F)$.

Clearly, if Conjecture P\ref{conj:physical_conj} is valid, then $\delta P_n$ must necessarily be $\le 0$ for all perturbations, including those considered in Eqs.~\eqref{eq:perturbed_bosonic_state} and \eqref{eq:perturbations} with  $\bm{v}$ being the eigenvector $\bm{v}_{\max}(F)$ corresponding to $\lambda_{max}(F)$, so we must have
\begin{equation}
\label{eq:ifP2wastrue}
\lambda_{max}(F) \le \perm{H} .
\end{equation}  
Since $\perm{H}$ is an eigenvalue of $F$ and thus, of course, $\perm{H} \le \lambda_{max}(F)$,
Eq.~\eqref{eq:ifP2wastrue} logically implies  Conjecture~M\ref{conj:bapatSunder 2}, \textit{i.e.}, $\lambda_{max}(F) = \perm{H}$. We must note, however, that Conjecture~P\ref{conj:physical_conj} implies but is not equivalent to  Conjecture~M\ref{conj:bapatSunder 2}. First, this is because we have restricted our argument to the perturbations of Eq.~\eqref{eq:perturbed_bosonic_state} with condition \eqref{eq:perturbations}, whereas others might violate Conjecture~P\ref{conj:physical_conj} even if Conjecture~M\ref{conj:bapatSunder 2} holds. Second, even for these specific perturbations, Conjecture~M\ref{conj:bapatSunder 2} is not a sufficient condition for the validity of Conjecture~P\ref{conj:physical_conj} as this would also require that the next contributing term in this series expansion is negative (in case the second-order term vanishes).  Thus, the logical implication P\ref{conj:physical_conj}$\,\Rightarrow\,$M\ref{conj:bapatSunder 2} holds whereas M\ref{conj:bapatSunder 2}$\,\Rightarrow\,$P\ref{conj:physical_conj} does not hold.

The contrapositive of implication P\ref{conj:physical_conj}$\,\Rightarrow\,$M\ref{conj:bapatSunder 2} is that, if Conjecture~M\ref{conj:bapatSunder 2} is false, then Conjecture P\ref{conj:physical_conj} must be false too. Specifically, $\delta P_n$ may become $>0$ as soon as 
\begin{equation}\label{eq:conj1_bunching}
    \lambda_{max}(F)>\perm{H}, 
\end{equation}
that is, if the matrix $H$ is a counterexample to Conjecture~M\ref{conj:bapatSunder 2}. We will exploit this idea in the next Section to build an explicit physical counterexample violating Conjecture P\ref{conj:physical_conj}, in which multimode bunching probabilities can unexpectedly be enhanced if a well-chosen perturbation is applied to the state of indistinguishable photons (making them only slightly distinguishable).  

%the local version of the generalized bunching conjecture 

\section{Enhancing bunching with nearly indistinguishable bosons}\label{sec:counterexamples}

In 2018, Drury found a $8\times 8$ positive semidefinite matrix $A$ that demonstrates that Conjecture M\ref{conj:bapatSunder 2} is false \cite{drury2016}. The matrix is defined as $A=M^{\dagger}M$, with 
\begin{align}
\label{eq:drurymatrix}
&M=M_R+ i M_I, \nonumber\\
& M_R= \begin{pmatrix}
-7& 9& -6 &3& 7 &4& 0& 5\\
4 &1 &-8 &-7& 1& 1& 8& 1\\
 \end{pmatrix}, \nonumber\\
& M_I= \begin{pmatrix}
4& -3& 2 &4& 6 &-4& 1& -8\\
-5 &4 &-2 &4&-4& -8& -6& -3\\
 \end{pmatrix}. 
 \end{align}
By calculating the matrix $F$  as defined in Conjecture M\ref{conj:bapatSunder 2} associated with Drury's matrix $A$, it can be seen that $\lambda_{max}(F)/\perm{A}\approx 1.017$, that is, this example violates the conjecture by about  $1.7\%$. 

This mathematical counterexample can be used to construct a physical situation where bunching can be enhanced by perturbing the state of indistinguishable photons. It was already noticed in Ref.~\cite{bosonbunching} that, given any positive semidefinite matrix $A$, we can find an interferometer $U$ and a subset $\mathcal{K}$ such that the corresponding  matrix $H$ in Eq.~\eqref{eq:H_Shchesnovich} is a rescaled version of $A$. In a nutshell, the idea is to embed $\sqrt{\gamma} M$ in the top left corner of $U$. The value of $\gamma$ is chosen such that $|| \gamma M^\dagger M ||= 1$.

To embed $\sqrt{\gamma} M$ into $U$ for the matrix $M$ defined in Eq.~\eqref{eq:drurymatrix}, it is enough to choose $U$ to be a $10\times 10$ matrix.  First, we add two columns to the matrix $\sqrt{\gamma} M$ choosing these columns such that the newly constructed $2\times 10$ matrix has orthonormal rows which are normalized (see also Lemma 29 from Ref.~\cite{aaronson2011computational}). Moreover, the lower part of $U$ (a $8\times 10$ matrix) can be constructed by generating 8 orthonormal row vectors that are also orthogonal to the first two rows. Such a $U$ could, in principle, be implemented with a circuit of 36 beam splitters, using for example the decomposition of Clements \textit{et al.} \cite{Clements}. 

%All the data for the physical implementation is available on github, see Section ``Code availability".

The optical experiment leading to a counterexample of Conjecture P\ref{conj:physical_conj} involves thus an input state of 8 photons, as the number of photons is given by the size of the matrix $H$, such that each photon enters one of the first 8 modes of a 10-mode interferometer. Since the rank of $H$ is 2, we consider the bunching probability of the 8~photons into 2 output modes, which can be chosen to be the first 2 modes given the freedom in how $U$ is constructed. We will see that this bunching probability can gradually increase by applying a suitable perturbation to the internal photons' state.

As demonstrated in Sec.~\ref{sec:second-order boson bunching}, the upper bound on the variation of the bunching probability $\delta P_n$ from Eq.~\eqref{eq:upper_bound} is attained by perturbing the internal wavefunction of indistinguishable photons in a two-dimensional space spanned by two orthogonal states $\{\ket{\phi_0}, \ket{\phi_1}\}$. Naturally, we refer to these states as  $\ket{\phi_0}\equiv \ket{H}$ and  $\ket{\phi_1}\equiv\ket{V}$ since polarization provides a natural two-dimensional degree of freedom of the photon that can be manipulated, for instance, using waveplates. Of course, any two-dimensional degree of freedom could be used, such as a twin time-bins encoding. We consider the initial situation where all photons have an horizontal polarization $\ket{H}$ and define the perturbed internal states as 
\begin{equation}
\label{eq:perturbed_bosonic_state_HV}
     |\phi_{i}\rangle=
     \frac{1}{\alpha_{i}}
     (\ket{H} + \epsilon~ {(\bm{v}_{\max})}_{i} \ket{V}),~
\end{equation}
with $i\in [8]$, where $\alpha_i$ are normalization constants. Here, ${(\bm{v}_{\max})}_{i}$ is the $i$th component of the normalized eigenvector corresponding to the largest eigenvalue of the $F$ matrix associated with the $H$ matrix constructed itself from Drury's counterexample (as explained above).  From  Eq.~\eqref{eq:maximum_bunching}, we see that the ratio between the two-mode bunching probability for the slightly distinguishable perturbed state and its counterpart for perfecrtly indistinguishable photons is given by 
\begin{align}\label{eq:violation_ratio}
    R(\epsilon)&=\frac{ \perm{H\odot \Tilde{S}}}{\perm{H}},\nonumber \\
    &\approx  1+ \epsilon^2 \left( \frac{\lambda_{\max}(F)}{\perm{H}}-1\right), \nonumber \\
    & \approx 1+  0.017 \,\epsilon^2, 
\end{align}
where we have neglected terms of $O(\epsilon^4)$. 

\begin{figure}[t]
    \centering
    \includegraphics[width = 0.5 \textwidth]{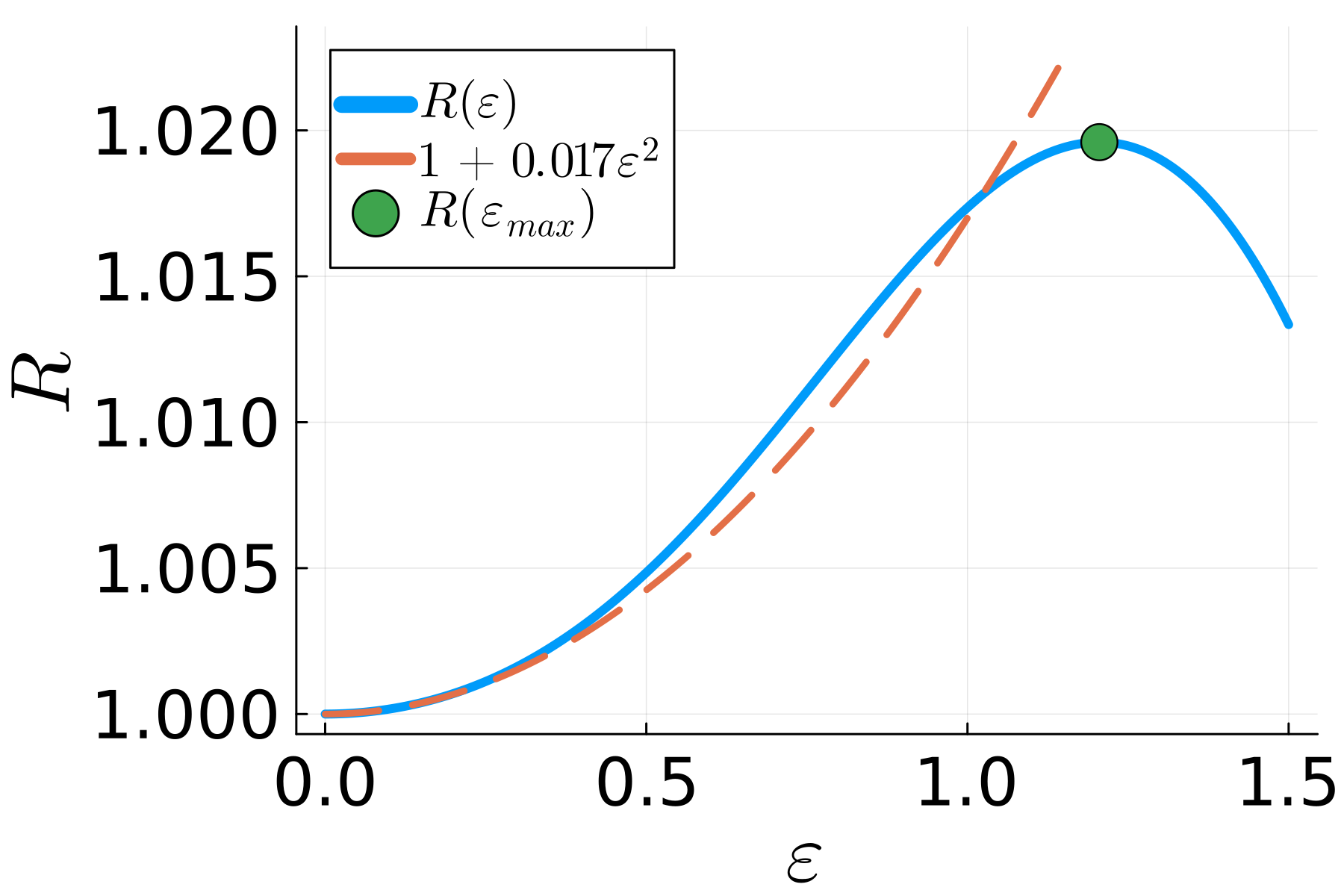}
    \caption{
    Violation ratio $R$ (solid line) as defined in Eq.~\eqref{eq:violation_ratio}, quantifying the relative enhancement of the bunching probability as a function of the perturbation strength $\epsilon$.  The perturbative approximation to second order $\epsilon^2$ is also plotted for comparison (dashed line). We observe that $R(\epsilon)$ exceeds $1$ in the neighborhood of indistinguishable particles, implying a violation of Conjecture P\ref{conj:physical_conj} (and hence of P\ref{conj:generalizedBunching} too).}
    \label{fig:ViolationRatio}
\end{figure}

Although our perturbative analysis is only accurate for small values of the perturbation strength $\epsilon$, we can exhibit a larger range of this parameter using numerics.  In Fig.~\ref{fig:ViolationRatio}, we show that the violation ratio $R(\epsilon)$ peaks for a value $\epsilon_{\max}\approx 1.2$, leading to a bunching enhancement of about $2\%$. Although a small effect, it is enough to demonstrate that the state of indistinguishable photons does not always locally maximize bunching probabilities in the space of possible (separable) input states.

\begin{figure}[t]
    \centering
    \includegraphics[width = 0.5 \textwidth]{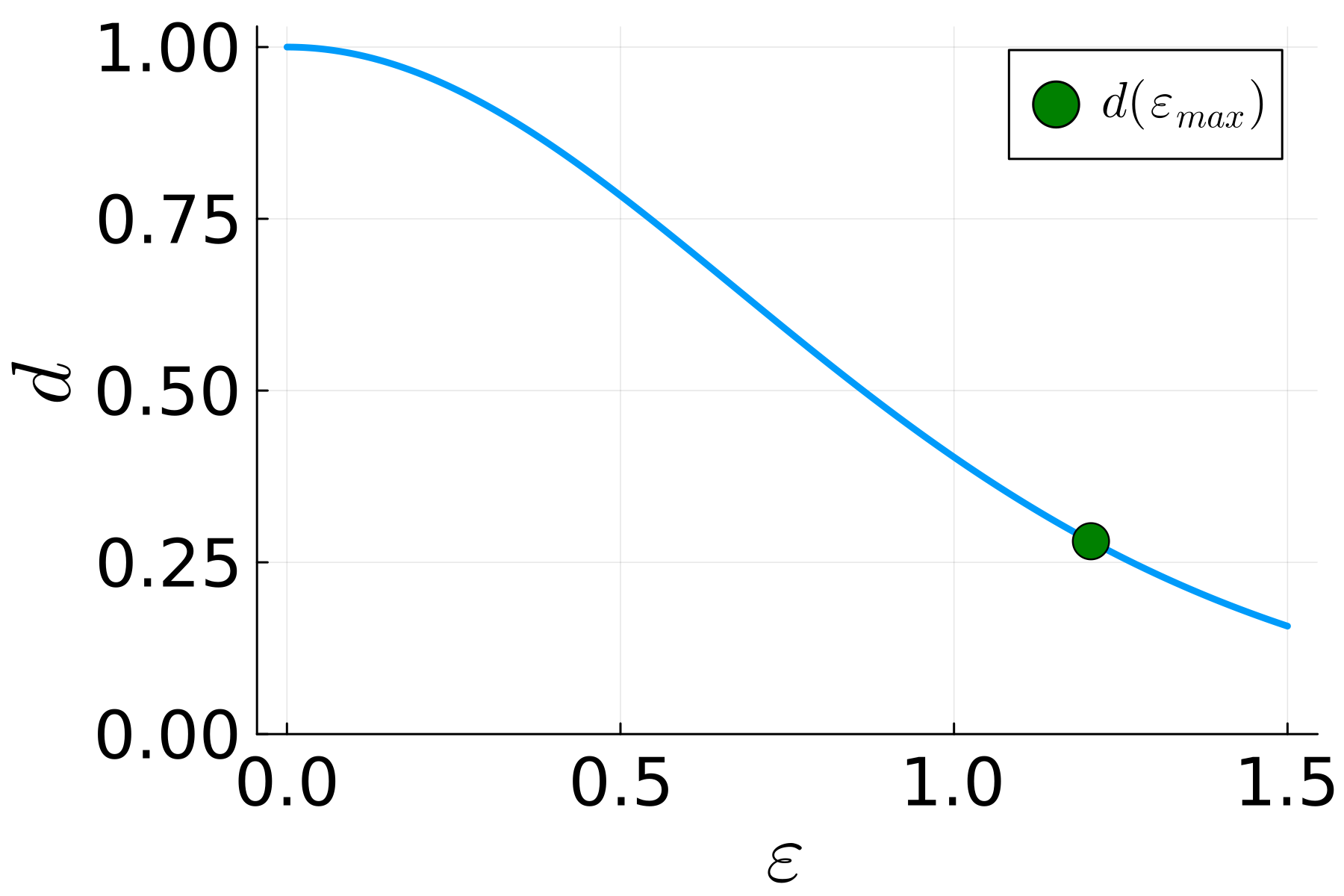}
    \caption{
     Measure of photon indistinguishability  $d(\tilde{S}(\epsilon))$ [Eq.~\eqref{eq:indistinguishability}] as a function of the perturbation strength $\epsilon$. This measure decreases monotonically with $\epsilon$, in contrast with the multimode bunching probability $P_n(\tilde{S}(\epsilon))$ which increases for $0\leq \epsilon\leq \epsilon_{max}$ before returning to the awaited decreasing behavior for  $\epsilon > \epsilon_{max}$. 
     }
    \label{fig:indistinguishability}
\end{figure}

It is also interesting to examine the evolution of photon indistinguishability as a function of $\epsilon$, according to the measure from Eq.~\eqref{eq:indistinguishability}, namely
\begin{equation}
    d(\tilde{S}(\epsilon))=\frac{\perm{\tilde{S}(\epsilon)}}{n!},
\end{equation}
for the input state of Eq.~\eqref{eq:perturbed_bosonic_state_HV}. We plot this quantity in Fig.~\ref{fig:indistinguishability}, showing that it always decreases monotonically in the range of $\epsilon$ that we explored. This automatically implies that 
\textit{single-mode} bunching probabilities decrease as a function of $\epsilon$, as expected since they are directly proportional to this indistinguishability measure \cite{tichy2015_partial_distinguishability}. Comparing Fig.~\ref{fig:ViolationRatio} with Fig.~\ref{fig:indistinguishability} corroborates
the findings of Ref.~\cite{bosonbunching}, namely that there exist instances where  multimode bunching behaves in an anomalous way with respect to $d(S)$ and hence single-mode bunching.

The peak value of the multimode bunching probability at $\epsilon_{\max}$ is $P_{n}(S(\epsilon_{\max}))\approx1.353\times 10^{-3}$, in comparison with $P_{n}(S(0))\approx1.327\times 10^{-3}$ for indistinguishable particles, making it very challenging to measure this effect with current technology. We leave as an open question whether higher violations could be observed, which would be important to facilitate the experimental observation of this phenomenon. Note also that, at the peak of $P_n$, the value of the indistinguishability measure is  $d(S(\epsilon_{\max})) \approx 0.281$, which is significantly higher than the observed value of    $d(S)\approx 8.93\times 10^{-3}$ for the 7-photon counterexample to Conjecture~P\ref{conj:generalizedBunching} \cite{bosonbunching}. The counterexample to Conjecture P\ref{conj:physical_conj} exhibited here is indeed much closer to the state of perfectly indistinguishable photons. We stress that the matrices $H$ from the counterexamples to Conjecture M\ref{conj:bapatSunder 1}/P\ref{conj:generalizedBunching} used in Ref.~\cite{bosonbunching} are \emph{not} counterexamples to Conjecture M\ref{conj:bapatSunder 2}, implying that they cannot be used to find violations of the (weaker) local version of the multimode bunching conjecture using our construction. This is consistent with the fact that the logical implication from M\ref{conj:bapatSunder 1} to  M\ref{conj:bapatSunder 2} is in one direction only (falsifying M\ref{conj:bapatSunder 1} does not imply falsifying  M\ref{conj:bapatSunder 2}). 

%In Table~\ref{tab:value_epsilon_max} we give the values of this measure evaluated at $\epsilon_{\max}$ as well as the violation ratio $R_n{(\epsilon_{\max}})$. As a comparison, for the 7-photon counterexample of the generalized bunching conjecture from Ref.~\cite{bosonbunching}, the indistinguishability measure $d(S)\approx 8.93\times 10^{-3}$, which is significantly lower than the values presented in Fig.~\ref{fig:indistinguishability}. 

\section{Decreasing bunching via small perturbations }
\label{sec:Further analysis of multimode boson bunching}

So far, we have established that even a small perturbation to indistinguishable photons can surprisingly increase multimode bunching probabilities in some well-chosen scenarios. To have a better understanding of the sensitivity of bunching probabilities to small perturbations, we also investigate here the opposite question: starting from the state of perfectly indistinguishable photons, what is the perturbation that leads to the strongest \emph{decrease} of both single-mode and multimode bunching probabilities? 

Naively, one could expect that such a perturbation should move the photons' internal state towards the state of fully distinguishable photons corresponding to mutually orthogonal internal states, that is
\begin{equation}
\label{eq:interpolation_model}
     |\Tilde{\phi_{i}}\rangle=
     \sqrt{1-\frac{\epsilon^2}{n}} \ket{\phi_{0}} + \frac{\epsilon}{\sqrt{n}} \ket{\eta_{i}},
\end{equation}
where all states $\ket{\eta_i}$ are orthogonal to $\ket{\phi_{0}}$ and also mutually orthogonal. Hence, the Gram matrix associated with the perturbation is the identity matrix $(\Delta_{ij} = \braket{\eta_i}{\eta_j}=\delta_{i,j})$. Equation~\eqref{eq:interpolation_model} corresponds to the most common model of partial distinguishability, which simply interpolates linearly between indistinguishable ($\epsilon=0$) and distinguishable ($\epsilon=\sqrt{n}$) particles (see, e.g., Ref.~\cite{tichy2015_partial_distinguishability}). The perturbation of Eq.~\eqref{eq:interpolation_model} takes the same form as that of Eq.~\eqref{eq:perturbed_bosonic_state} if we choose $v_i=1/\sqrt{n}, ~\forall i$, and neglect terms of $O(\epsilon^3)$. Following a similar treatment to that of Sec.~\ref{sec:second-order boson bunching}, namely using Eq.~\eqref{eq:second_order_bunching_with_vi}, we can thus express the variation of the bunching probability as 
\begin{equation}\label{eq:deltaP_interpolation}
    \delta P_n \approx \epsilon^2 \left(\frac{\text{tr}(F)}{n}- \perm{H}\right).
\end{equation}

For comparison, let us consider instead a perturbation similar to the one used to reach the upper bound of $\delta P_n$, namely moving each photon's internal state inside the two-dimensional space spanned by the orthogonal states $\{\ket{\phi_0}, \ket{\phi_1}\}$. In order to minimize the bunching probability, it is natural to consider the perturbed states
\begin{align}
\label{eq:perturbed_bosonic_state_vmin}
     |\Tilde{\phi_{i}}\rangle=
   \sqrt{1-\epsilon^2|{(\bm{v}_{\min})}_{i}|^2} \ket{\phi_0} + \epsilon~ {(\bm{v}_{\min})}_{i} \ket{\phi_1}, 
\end{align}
where $\bm{v}_{min}$ is the eigenvector of $F$ corresponding to its minimum eigenvalue $\lambda_{min}(F)$. Thus, the variation of the bunching probability is again described by Eq.~\eqref{eq:maximum_bunching}, but yields this time
\begin{equation}\label{eq:deltaP_vmin}
     \delta P_n \approx \epsilon^2 \Big(\lambda_{min}(F)- \perm{H}\Big). 
\end{equation}
 In Appendix~\ref{sec:Minimization of multimode bunching}, it is shown that Eq.~\eqref{eq:deltaP_vmin} is a lower bound to $\delta P_n$ for all perturbations considered in Eqs.~\eqref{eq:perturbed_bosonic_state} and \eqref{eq:perturbations} [this is proven using a similar approach as for the upper bound in Eq.~\eqref{eq:upper_bound}]. 
%described in Sec.~\ref{sec:second-order boson bunching},
Consequently, the naive perturbation towards the state of fully distinguishable particles described by Eq.~\eqref{eq:interpolation_model} cannot  decrease the bunching probability further than the one described by Eq.~\eqref{eq:perturbed_bosonic_state_vmin}, even though the latter only changes the photons' internal states in a two-dimensional subspace (which is not enough to encompass $n$ fully distinguishable photons). Actually, by comparing Eq.~\eqref{eq:deltaP_interpolation} with Eq.~\eqref{eq:deltaP_vmin}, it is evident that the naive perturbation of Eq.~\eqref{eq:interpolation_model} effects a lower decrease than the perturbation of 
Eq.~\eqref{eq:perturbed_bosonic_state_vmin} whenever $\lambda_{min}(F)< \text{tr}(F)/n $, which occurs as soon as $F$ admits several eigenvalues.

Although this is still another surprising phenomenon, it does not lead to an anomalous mismatch between single-mode and multimode bunching (unlike the violation of Conjecture~P\ref{conj:generalizedBunching} ). Indeed, $\lambda_{min}(F)< \text{tr}(F)/n $ is very possible even for single-mode bunching ($|\mathcal{K}|=1$), whereas $\lambda_{max}(F) > \perm{H}$ may only take place for $|\mathcal{K}|>1$.
Accordingly, we observe that the perturbed state described by Eq.~\eqref{eq:interpolation_model} is not the one that maximizes the local decrease of single-mode bunching starting from perfectly indistinguishable photons (it does not lead to the maximum decrease of $d(\epsilon)$ around $\epsilon=0$). 

The perturbation that decreases single-mode bunching the most corresponds to Eq.~\eqref{eq:perturbed_bosonic_state_vmin}, leading to
\begin{equation}
     \delta P_n \approx - \epsilon^2  \,\perm{H} 
\end{equation}
as it appears that $\lambda_{min}(F)=0$ in this case. Indeed, the matrix $H$ is a rank-one matrix for single-mode bunching, which implies that $F$ is rank-one too; hence, its minimum eigenvalue is zero (remember that $F$ is positive semidefinite). An alternative description of this situation is provided in Appendix~\ref{sec:Minimization of single-mode bunching}, based on a second-order perturbation analysis of $\perm{S}$ instead of $\perm{H \odot S}$, exploiting the fact that $\perm{H \odot S}$ factorizes. It appears that the  perturbations that decrease single-mode bunching the most are those for which the matrix $\Delta$ has a null eigenvalue and $\bm{v}$ is the eigenvector associated with this null eigenvalue. This situation can arise as soon as the dimension of the perturbation space is smaller than the number of photons. This confirms that in order to locally decrease single-mode and multimode boson bunching, the best perturbations are those with a small internal space dimension.

%Contrary to what might be expected

\section{Conclusion}
\label{eq:Conclusion}
%need to discuss 
%-what you showed in your article
%- what are the implications of your work
%- what new questions does it rais

In this work, we have built on mathematical conjectures regarding matrix permanents whose physical interpretation relates to boson bunching. Our central result is the identification of an optical circuit and a perturbation which, when applied to the ideal state of perfectly indistinguishable photons, enhances (instead in decreasing) the probability for all photons to bunch into two output modes. Although it had recently been discovered that indistinguishable bosons do not necessarily always maximize multimode bunching probabilities~\cite{bosonbunching}, it remained quite plausible that this ideal situation describes at least a local maximum. However, the counterexample revealed in the present work implies that indistinguishable bosons do not even locally maximize the multimode bunching probability, disclosing another surprising facet of anomalous boson bunching.

By randomly generating unitary matrices U based on the Haar measure, we have found other matrices $H$ that contradict Conjecture~M\ref{conj:bapatSunder 2}. This could be done by choosing for perturbation the appropriate eigenvector of the matrix $F$ associated with $H$. Note, however, that all numerically found counterexamples corresponded to an optical setup with at least 8 photons. It is, however, still worth looking for simpler scenarios (involving less than 8 photons in a smaller interferometer) where such an anomalous bunching effect could appear more pronounced, bringing it closer to a practical experimental implementation and potentially improving our understanding of this phenomenon.

In fact, it would be interesting to identify the physical mechanism responsible for the enhancement of bunching with nearly indistinguishable bosons, as could be done for the effect found in Ref.~\cite{bosonbunching}. Unfortunately, the mathematical counterexample of Conjecture~M\ref{conj:bapatSunder 2} exploited here seems to be less structured than the counterexample of Conjecture~M\ref{conj:bapatSunder 1} exploited in Ref.~\cite{bosonbunching}, and consequently so does its physical realization as a linear optical experiment. The underlying mechanism, therefore, remains an open question.

Along the same line, we have unveiled another counterintuitive behavior of boson bunching with respect to particle distinguishability. We have shown that some specific perturbation to the photonic internal state in a two-dimensional space is the one that maximally decreases bunching probabilities near indistinguishable photons when compared to all possible perturbations (possibly involving many degrees of freedom). Although this effect is visible with single-mode as well as multimode bunching (hence, it is not anomalous), it nevertheless goes against our intuition as it implies that a perturbation going towards fully distinguishable photons (for example, with all photons occupying non-overlapping time bins) is actually \textit{not} the path that maximizes the decrease of bunching probabilities.

As a possible direction for further investigation, it is worth recalling that an even more general form of the multimode bunching conjecture has been formulated which considers arbitrary -- possibly entangled -- input states and which can be related to a different conjecture about the permanent of positive semidefinite matrices (the so-called ``permanent-on-top'' conjecture) \cite{shchesnovich2016universality}. A numerical counterexample to this conjecture was found involving 5 photons bunching into 2 modes \cite{shchesnovich2016permanent}.
This counterexample does not automatically imply the existence of a counterexample to Conjecture~P\ref{conj:generalizedBunching} (hence to Conjecture~P\ref{conj:physical_conj}), because
the ``permanent-on-top'' conjecture implies Conjecture~P\ref{conj:generalizedBunching} but the converse does not hold. It would be nice to reach a better understanding of the mechanism behind the violation of the ``permanent-on-top'' conjecture, as this could help finding an explicit physical realization of this violation via an optical circuit and an appropriate input state (presumably even simpler than those contradicting Conjecture~P\ref{conj:generalizedBunching} or P2 since this conjecture is stronger).

As a final note and despite the existence of several kinds of anomalous bunching phenomena as found in this work as well as Refs.~\cite{shchesnovich2016permanent,bosonbunching}, we believe that multimode bunching probabilities can be useful quantities to measure as experimental witnesses of particle indistinguishability in many practical scenarios. Hence, we hope our work will inspire new research toward clarifying in which physical situations the observation of multimode bunching qualifies as a faithful indistinguishability witness. 

%\vspace{-0.5cm}
\section*{Code availability}
\label{sec:code_availability}
Numerical simulations were performed using the package \textsc{Permanents.jl} \cite{seron2023permanentsjl,seron2022bosonsampling}. The source code generating the figures and data of this paper is available on \href{https://github.com/benoitseron/generalized_bunching}{github}.

%\vspace{-0.4cm}
\section*{Note added}
A side result of this paper is the proof of a logical implication between two different conjectures from Bapat and Sunder (M\ref{conj:bapatSunder 1}~$\Rightarrow$~M\ref{conj:bapatSunder 2}), which then connects different properties of the permanent of positive semidefinite matrices. We believe that this physics-inspired result may be of independent interest to the mathematics literature as the investigation of conjectures on the permanents of positive semidefinite matrices remains an active research area, due in particular to its connection with the long-standing Lieb's permanental dominance conjecture~\cite{Wanless}. We give a variant proof of this implication in Ref.~\cite{math-paper} without using the quantum mechanics jargon, so it is accessible to a broader mathematics community.

%We also find out that the perturbation that maximally decreases the bunching probability is not the one that takes the perfectly indistinguishable state towards a fully distinguishable state, as could naively be expected. 

%as our intuition dictates. 

%\vspace{-0.5cm}
\section*{Acknowledgments}
We would like to thank Andreas Buchleitner for useful suggestions and Ian Wanless for useful correspondence. L.P. and N.J.C. acknowledge support from the Fonds de la Recherche Scientifique – FNRS and from the European Union under project ShoQC within ERA-NET Cofund in Quantum Technologies (QuantERA) program. L.P. is a FRIA grantee of the Fonds de la Recherche Scientifique – FNRS. 
B.S. acknowledges funding from Fonds de la Recherche Scientifique – FNRS and Georg H. Endress Foundation. 
L.N. acknowledges funding from FCT-Fundação para a Ciência
e a Tecnologia (Portugal) via the Project No.
CEECINST/00062/2018, and from the European Union's Horizon Europe research and innovation program under EPIQUE Project GA No. 101135288. N.J.C. also acknowledges funding from the European Union’s Horizon 2020 research and innovation programme under Marie Skłodowska-Curie grant agreement no. 956071.
\bibliographystyle{unsrt}
\bibliography{main}

\clearpage
\onecolumngrid

\section{Counterexamples to Conjecture~P\ref{conj:generalizedBunching} with $|\mathcal{K}|\geq n$}
\label{sec:appendix_counterexamples}
\begin{figure}[h]
    \centering
    \includegraphics[width = 0.6\textwidth]{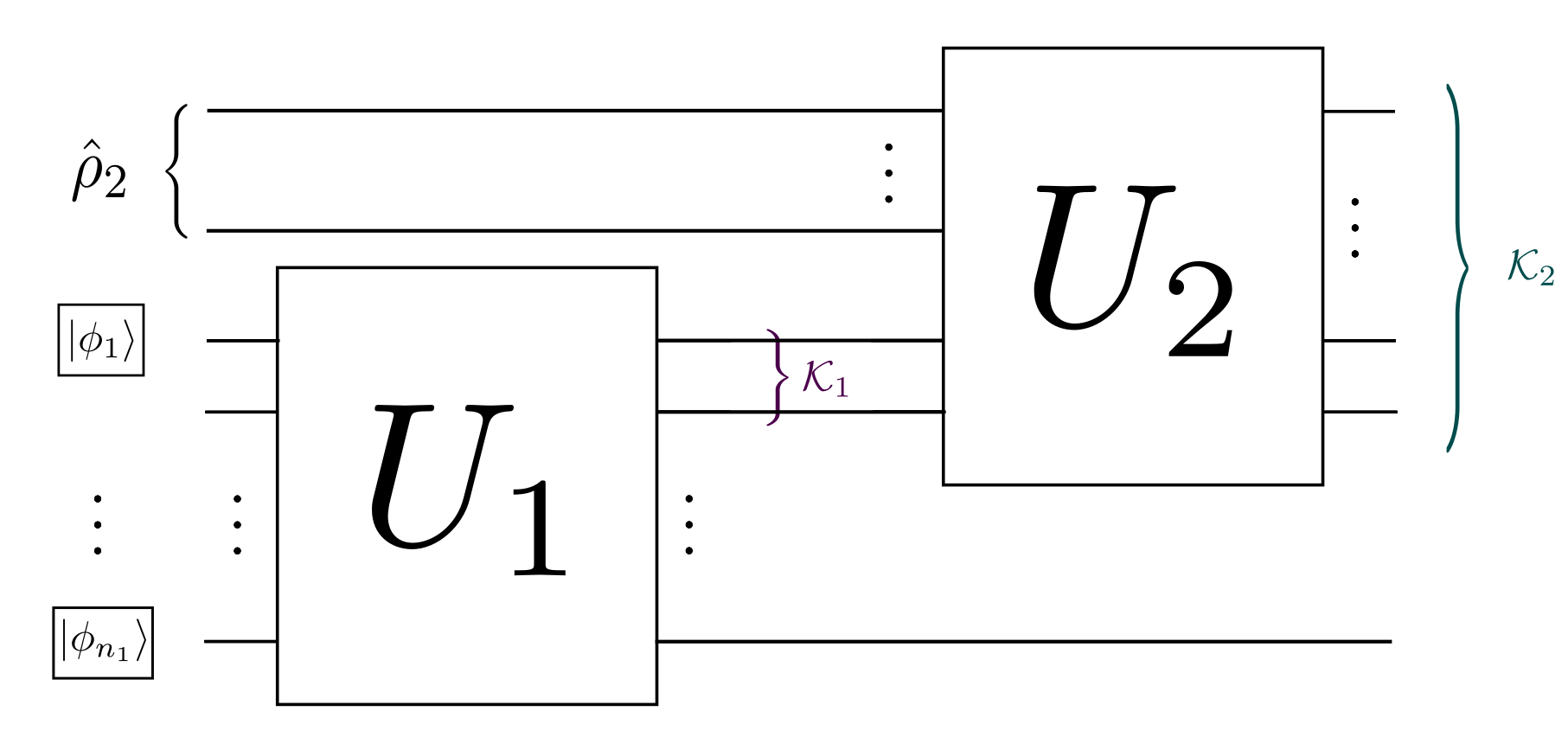}
    \caption{Counterexamples to Conjecture~P\ref{conj:generalizedBunching} with $|\mathcal{K}|<n$ can be trivially extended to counterexamples with $|\mathcal{K}|\geq n$ by encompassing them into a larger circuit.}
    \label{fig:more_modes_than_photons}
\end{figure}

In this appendix, we show how counterexamples to P\ref{conj:bapatSunder 1} with $|\mathcal{K}|<n$ can be trivially extended to counter-examples with $|\mathcal{K}|\geq n$. 
Consider a circuit $U_1$ with $n_1$ inpinging photons who violate P\ref{conj:generalizedBunching} for some choice of internal states $\ket{\phi_1},..., \ket{\phi_{n_1}}$ and of subset $\mathcal{K}_1$ (see Fig.~\ref{fig:more_modes_than_photons}).
To construct a counterexample with $|\mathcal{K}|\geq n$, we use the output modes of $U_1$ defined by $\mathcal{K}_1$ as inputs of a new network, $U_2$. The remaining modes of $U_2$ are populated by an input state $\hat{\rho}_2$, with $n_2$ single photons in a classically correlated or uncorrelated state. 
Considering all output modes of $U_2$ as a subset $\mathcal{K}_2$, we obtain a violation for the entire circuit ($U_1, U_2$ combined as shown in Fig.~\ref{fig:more_modes_than_photons}). As long as $n = n_1+n_2 \leq |\mathcal{K}_2|$ (for instance, by choosing $\hat{\rho}_2$ as the vacuum), we obtain a counterexample with less photons than the size of the subset ($\mathcal{K} = \mathcal{K}_2$). We also note that, even though in this construction the total interferometer $U$ does not couple all input modes to all output modes, it has been previously shown that violations to Conjecture~P\ref{conj:generalizedBunching} exhibit a certain robustness to perturbations of the matrix elements of $U$ \cite{bosonbunching}. Hence, it should be possible to find counterexamples not only with $|\mathcal{K}|>n$ but also where all matrix elements describing the linear interferometer are non-zero.   

%\textcolor{blue}{The same question about the size of the subset $\mathcal{K}$ arises in the context of the ``permanent-on-top'' conjecture (i.e., with entangled input states), which was originally formulated with the restriction $|\mathcal{K}|\geq n$ \cite{shchesnovich2016universality}.  An even more general form of the conjecture can be stated too, with an \emph{arbitrary} subset size $|\mathcal{K}|$ possibly smaller than $n$; this  was disproved by the counterexample found in Ref. \cite{shchesnovich2016permanent}%with $n=5$ and $|\mathcal{K}|=2$. }

\section{Positive semidefiniteness of matrix $F$}

\label{appendix:F_psd}

Let $A$ be a $n\times n$ positive semidefinite Hermitian matrix. The $n\times n$ matrix $F$ is
defined as (see Conjecture M\ref{conj:bapatSunder 2})
\begin{equation}
    F_{i,j}=A_{i,j} \, \perm{A(i,j)},
\label{eq:def_of_F}
\end{equation} 
where $A(i, j)$ denotes the $(n - 1)\times (n - 1)$ submatrix of $A$ obtained by deleting the $i$th row and $j$th column of $A$. Of course, Eq.~\eqref{eq:def_of_F} is equivalent to Eq.~\eqref{eq:F_definition} where $A$ has been replaced by $H$.
Let us prove that the matrix $F$ is positive semidefinite. We write $F_{i,j} = (A \odot B)_{i,j}$,
with $B_{i,j} = \perm{A(i,j)}$, and note that it suffices to prove that the matrix $B$ is positive semidefinite. Indeed, since $A$ is positive definite, the positive semidefiniteness of $B$ implies that $A \odot B $ is positive semidefinite as a consequence of Schur's Theorem (see Theorem 7.21 on page 234 of Ref. \cite{zhang2011matrix}).

We recall a formalism due to Minc (see Section 2.2. of Ref. \cite{Minc}). Consider a matrix $A_{ij} = \braket{u_i}{v_j}$ whose generating vectors $\bm{u_1}, \dots, \bm{u_n},\bm{v_1}, \dots, \bm{v_n}$ belong in a n-dimensional unitary space $V$. We write the tensor products of the generating vectors as $\bm{u_1} \otimes ... \otimes \bm{u_n} \in V^n$. We also introduce the symmetric product 
\begin{equation}
\bm{u_1} * ... * \bm{u_n}= T_n(\bm{u_1} \otimes ... \otimes \bm{u_n}),
\end{equation}
where  we have defined 
the completely symmetric operator 
\begin{equation}
T_n = \frac{1}{n!} \sum _{\sigma \in S_n} P(\sigma),
\end{equation}
with $P(\sigma)$ being the permutation operator on $V^n$ corresponding to the permutation $\sigma$, namely
\begin{equation}
P(\sigma)\left(\bm{u}_{1} \otimes \cdots \otimes \bm{u}_{n}\right)=\bm{u_{\sigma^{-1}(1)}} \otimes \cdots \otimes \bm{u_{\sigma^{-1}(n)}}.
\end{equation}
It was shown by Minc that the permanent of $A$ can be expressed from the generating vectors of $A$ via
\begin{equation}
\perm{A} = n! \, (\bm{u_1} * ... * \bm{u_n}, \bm{v_1} * ... * \bm{v_n}),
\end{equation}
where the scalar product on $V^n$ is defined as
\begin{equation}
(\bm{u_1} \otimes ... \otimes \bm{u_n}, \bm{v_1} \otimes ... \otimes \bm{v_n}) = \prod _{i=1}^n \braket{u_i}{v_i} .
\end{equation}
Apply this formalism to our case where $A$ can be written as $A_{ij} = \braket{u_i}{u_j}$ since it is positive semidefinite, we have
\begin{equation}
B_{i,j}=  \perm{A(i,j)} = (n-1)! \, (\bm{u_1} * ... * \bm{\hat{u}_i}*  ... * \bm{u_n}, \bm{u_1} * ... * \bm{\hat{u}_j} * ... * \bm{u_n}),
\end{equation}
where the hat in $\hat{u}_i$ indicates that this vector is excluded from the symmetric product. Given this decomposition, we see that the matrix $B$ is itself positive semidefinite since its elements are defined from scalar products (just as $A$).

\section{Minc's formula for the permanent of a sum}
\label{sec:Minc'sformula}
Minc's formula \cite{caianiello1959regularization, Minc} makes it possible to expand the permanent of the sum of two matrices, $A$ and $B$, as follows
\begin{equation}
\label{eq:minc}
    \perm{A+B} = 
    \sum_{r=0}^n \, \sum_{\alpha,\beta \in Q_{r,n}} \perm{A[\alpha,\beta]}\perm{B(\alpha,\beta)},
\end{equation}
where $Q_{r,n}$ is the set of increasing sequences defined as
 \begin{equation}
Q_{r, n}=\left\{\left( u_{1}, \ldots,  u_{r}\right) \in \Gamma_{r, n} \mid 1 \leq  u_{1}<\cdots< u_{r} \leq n\right\},
\end{equation}
with $\Gamma_{r, n}$ being the set of all $n^{r}$ sequences $ u=\left( u_{1}, \ldots,  u_{r}\right)$ of integers, with $1 \leq u_{i} \leq n, i=1, \ldots, n$.
In Eq.~\eqref{eq:minc}, $A[\alpha,\beta]$ represents the $r\times r$ matrix formed by selecting the rows and columns of matrix $A$ based on the sets $\alpha$ and $\beta$. Conversely, $A(\alpha,\beta)$ denotes the $(n-r)\times (n-r)$ matrix obtained after excluding the rows and columns specified by $\alpha$ and $\beta$ from $A$.

\section{Minimization of the bunching probability}
\label{sec:Minimization of multimode bunching}

In order to determine the specific  perturbation that minimizes the bunching probability near the state of indistinguishable photons, we look for a lower bound of $P_{n}(\Tilde{S})$ and demonstrate that this bound is reached. To achieve this, we start from Eq. \eqref{eq:before_upper_bound}, namely
\begin{equation}    \label{eq:before_lower_bound_appendix}
     \delta P_n=\epsilon^{2} \Bigl( \sum_{k=1}^{K}(\bm{v}\odot \bm{u}^{k})^{\dagger} F (\bm{v}\odot \bm{u}^{k})\nonumber
     - \perm{H}\Bigl) +O(\epsilon^4)
\end{equation}
and define $\lambda_{min}(F)$ as the minimum eigenvalue of $F$. We will also introduce the family of vector $\bm{w}^k$, which are eigenvectors of $F$ associated with $\lambda_{min}(F)$ such that $||\bm{w}^k||=||\bm{v}\odot \bm{u}^k||$. We have
\begin{equation}
    \sum_{k=1}^{K}(\bm{v}\odot \bm{u}^{k})^{\dagger} F (\bm{v}\odot \bm{u}^{k})  \geq   \sum_{k=1}^{K}(\bm{w}^k)^{\dagger} F (\bm{w}^k).
\end{equation}
Since $\bm{w}^k$ is an eigenvector of $F$ and by using Eq.~\eqref{eq:norm_v_o_u_equal_1}, we have

\begin{equation}
   \sum_{k=1}^{K}(\bm{w}^k)^{\dagger} F (\bm{w}^k) = \lambda_{min}(F)\sum_{k=1}^{K}{ ||\bm{w}^k||^2}=\lambda_{min}(F).
\end{equation}
The lower bound on $ \delta P_n$ is, then

\begin{align}  
    \label{eq:lower_bound_appendix}
    \delta P_n\geq \epsilon^2(\lambda_{min}(F)-\perm{H})+O(\epsilon^4).
\end{align}
As before, we note that this bound is reached in the case where the total Hilbert space has a dimension of $2$, and $\bm{v}$ is chosen to be the eigenvector of the matrix $F$ with the smallest eigenvalue. In this situation, the bunching is decreased by an amount of $ \epsilon^2 (\lambda_{\min}(F)-\perm{H})+O(\epsilon^4)$ with the perturbation. It is worth noting that the matrix $F$ is positive semidefinite, as shown in Appendix \ref{appendix:F_psd}, so $\lambda_{min}(F)$ is always non-negative and the decrease of $P_n$ can never be stronger than $-\perm{H}$.

\section{Minimization of single-mode bunching}
\label{sec:Minimization of single-mode bunching}
In this section, we particularize the minimization of multimode boson bunching minimization to the case of single-mode boson bunching, i.e., the situation where all bosons bunch in a single output mode. In this particular case, the probability of bunching in the output mode $k$ is given by Eq.~\eqref{eq:single-mode-factorization}, namely
\begin{equation}
P_{n} = \prod_{i=1}^n|U_{k,i}|^2 \,\perm{S} ,
\end{equation}
In contrast to the general situation, $P_{n}$ is proportional to $\perm{S}$. It is, therefore, possible to study the impact of partial distinguishability independently of the interferometer. Let us consider the case of nearly indistinguishable bosons described by Eq.~\eqref{eq:perturbed_bosonic_state} and Eq.~\eqref{eq:perturbations}. As previously, the distinguishability matrix is given by:
\begin{equation}
\Tilde{S}_{i,j}=1+\epsilon^{2}\Bigl(v_{i}^{*}v_{j}\Delta_{ i,j}-\frac{|v_{i}|^{2}}{2}-\frac{|v_{j}|^{2}}{2} \Bigl)+\mathcal{O}(\epsilon^{4})=\mathbb{E}_{i,j} +\epsilon^{2}X_{i,j} +\mathcal{O}(\epsilon^{4}),
\end{equation}
with $X_{i,j}=v_{i}^{*}v_{j}\Delta_{i,j}-\frac{|v_{i}|^{2}}{2}-\frac{|v_{j}|^{2}}{2}$. By applying the same steps as in Sec.~\ref{sec:second-order boson bunching}, we obtain
\begin{equation}
\perm{\Tilde{S}}=\perm{\mathbb{E}+\epsilon^2 X } =\perm{\mathbb{E}} + \sum_{i,j}^{n}(\epsilon^{2}X_{i,j})\, \perm{\mathbb{E}(i,j)}    +O(\epsilon^4),
\end{equation}
\begin{equation}
     \perm{\Tilde{S}}=n!+ \epsilon^{2}(n-1)!\sum_{i,j}^{n}X_{i,j}\,+O(\epsilon^4).
\end{equation}
Since the vector $\bm{v}$ is normalized, we have
\begin{equation}     \sum_{i,j}^{n}X_{i,j}=\sum_{i,j}^{n}  v_{i}^{*}\Delta_{i,j}v_{j} - n||v||^2=(\bm{v}^{\dagger}\Delta \bm{v}-n).
\end{equation}
Now let us find which perturbation applied to the indistinguishable particle locally minimizes the single-mode bunching. Since $\Delta$ is a Gram matrix, its eigenvalues are lower bounded by 0, then
\begin{equation}
\label{eq:lower_bound_single-mode}
    \perm{\Tilde{S}}\geq n! -\epsilon^{2} \, n! +O(\epsilon^4).
\end{equation}
For single-mode boson bunching, among the perturbations satisfying Eq.~\eqref{eq:perturbations}, the ones that saturate the lower bound of 
Eq.~\eqref{eq:lower_bound_single-mode} are those where, firstly, the matrix $\Delta_{i,j}=\braket{\eta_i}{\eta_j}$ has an eigenvalue of zero. Secondly, the vector $\bm{v}$ equals the eigenvector of $\Delta$ associated with this eigenvalue. These situations are frequent, as they can always occur when the internal state dimension of the photons is smaller than the number of photons.\\

Although this is not true for multimode boson bunching, one could expect that for single-mode boson bunching, a perturbation that interpolates linearly between indistinguishable and distinguishable particles reaches the lower bound of Eq.~\eqref{eq:lower_bound_single-mode}. However, we will show that this intuition needs to be corrected. The Gram matrix associated with the perturbation is $\Delta_{ij} =\delta_{i,j}$, which implies that 
\begin{equation}
\sum_{i,j}^{n}X_{i,j}=\bm{v}^{\dagger}\bm{v}-n=1-n
\end{equation}
\begin{equation}
    \perm{\Tilde{S}}= n!\Big(1 -\epsilon^{2}\frac{n-1}{n}\Big) +O(\epsilon^4)
\end{equation}
Contrary to intuition, a perturbation using this model of partial indistinguishability does not lead to the maximum possible decrease of single-mode bunching probabilities.

\end{document}